# Probing the Progenitor of High-z Short-duration GRB 201221D and its Possible Bulk Acceleration in Prompt Emission

Hao-Yu Yuan[1], Hou-Jun Lü[1], Ye Li[2], Bin-Bin Zhang[3], Hui Sun[4], Jared Rice[5], Jun Yang[3], and En-Wei Liang[1]
[1] Guangxi Key Laboratory for Relativistic Astrophysics, Department of Physics, Guangxi University, Nanning 530004, China; lhj@gxu.edu.cn
[2] Purple Mountain Observatory, Chinese Academy of Sciences, Nanjing 210008, China
[3] Key Laboratory of Modern Astronomy and Astrophysics (Nanjing University), Ministry of Education, Nanjing 210093, China
[4] Key Laboratory of Space Astronomy and Technology, National Astronomical Observatories, Chinese Academy of Sciences, Beijing 100101, China
[5] Department of Mathematics and Physical Science, Southwestern Adventist University, Keene, TX 76059, USA
*Received 2022 February 22; revised 2022 May 9; accepted 2022 May 16; published 2022 June 14*

## Abstract

The growing observed evidence shows that the long- and short-duration gamma-ray bursts (GRBs) originate from massive star core-collapse and the merger of compact stars, respectively. GRB 201221D is a short-duration GRB lasting ∼0.1 s without extended emission at high redshift $z = 1.046$. By analyzing data observed with the Swift/BAT and Fermi/GBM, we find that a cutoff power-law model can adequately fit the spectrum with a soft $E_{\rm p} = 113^{+9}_{-7}$ keV, and isotropic energy $E_{\gamma,\rm iso} = 1.36^{+0.17}_{-0.14} \times 10^{51}$ erg. In order to reveal the possible physical origin of GRB 201221D, we adopted multi-wavelength criteria (e.g., Amati relation, $\varepsilon$-parameter, amplitude parameter, local event rate density, luminosity function, and properties of the host galaxy), and find that most of the observations of GRB 201221D favor a compact star merger origin. Moreover, we find that $\hat\alpha$ is larger than $2 + \hat\beta$ in the prompt emission phase which suggests that the emission region is possibly undergoing acceleration during the prompt emission phase with a Poynting-flux-dominated jet.

## 1. Introduction

Phenomenologically, cosmic gamma-ray bursts (GRBs) are classified into two categories with a division line at the observed duration of $T_{90} \sim 2$ s, named as "long" and "short" GRBs with $T_{90} > 2$ s and $T_{90} < 2$ s, respectively (Kouveliotou et al. 1993). The long and short GRBs seem to be consistent with massive star core-collapse and compact star mergers (Eichler et al. 1989; Woosley 1993). However, the duration $T_{90}$ of a GRB is energy-dependent and detector-sensitivity dependent (Qin et al. 2013). Searching for methods of classification to reveal the intrinsic origin have never stopped (Zhang 2018). Zhang (2006) proposed by using Type I (origin in compact star mergers) and Type II (origin in massive star core-collapse). Later, Zhang et al. (2009) used multi-wavelength criteria to diagnose the physical origin of a GRB. Lü et al. (2010) proposed defining the $\varepsilon$ parameter ($\varepsilon = E_{\gamma,\rm iso,52}/E_{\rm p,z,2}^{5/3}$) for GRBs with $z$ measurements by considering both the burst energy and the spectral properties in the rest frame, and found that the values of $\varepsilon$ have an approximately bimodal distribution corresponding to Type I and Type II origin (Zhang 2006). Moreover, Lü et al. (2014) suggested that some short GRBs with a lower amplitude parameter (defined as the ratio between peak flux and average background flux in the light curve of prompt emission) may be a "tip of the iceberg" for long GRBs due to observational effects.

Observationally, supernovae (SNe) are associated with some long GRBs (or without short GRBs; Galama et al. 1998; Hjorth et al. 2003; Soderberg et al. 2004; Campana et al. 2006) and the host galaxies of long (or short) GRBs are typically associated with irregular galaxies with intense (or little) star formation (Tanvir et al. 2005; Fruchter et al. 2006). These lines of observational evidence, as well as the joint detection of the gravitational-wave event GW170817 and the short GRB 170817A (Abbott et al. 2017; Goldstein et al. 2017; Savchenko et al. 2017; Zhang et al. 2018b), encourage people to believe that long and short GRBs are likely related to the deaths of massive stars (Type II) and the merger of two compact stellar objects (Type I), respectively (Eichler et al. 1989; Woosley 1993; Zhang 2006). However, some apparently long-duration (or short-duration high-z) GRBs have been suggested to originate in compact stars mergers (or massive star core-collapse). Two counterexamples are GRB 060614 (Gehrels et al. 2006), and GRB 090426 (Levesque et al. 2010; Xin et al. 2011). In particular, Zhang et al. (2021) recently discovered another peculiarly short-duration GRB 200826A which seems to originate in massive star core-collapse.

In either the death of massive star or merger of compact stars, the catastrophic event leaves behind a hyper-accreting black hole or a rapidly rotating highly magnetized neutron star (called a magnetar), which serves as the central engine of a collimated outflow (or jet) with a relativistic speed toward Earth (Usov 1992; Dai & Lu 1998b; Zhang & Mészáros 2001; Zhang 2011; Lü & Zhang 2014; Kumar & Zhang 2015; Lü et al. 2015; Chen et al. 2017). One basic question is what is the composition of the relativistic jets? There are two models widely discussed in the literature (Lei et al. 2013). One is the fireball model with a matter-dominated outflow which dissipates its kinetic energy in internal shocks or external shocks to produce the observed GRB emission (Rees & Meszaros 1992, 1994; Meszaros et al. 1993; Kobayashi et al. 1997). Within this model, the fireball has a rapid acceleration early on and can only reduce its kinetic energy at large radii from the central engine. The other one is Poynting-flux-dominated outflow. Within this scenario, the Poynting flux energy can be converted to kinetic energy (Drenkhahn & Spruit 2002; Komissarov et al. 2009), and then converted to particle energy and radiation via magnetic dissipation, such as reconnection, current instabilities, and internal collisions (Zhang & Yan 2011). In comparison to the fireball model, a Poynting-flux-dominated jet can undergo gradual acceleration in a large range of emission region (Gao & Zhang 2015; Uhm & Zhang 2015).

Traditionally, it is assumed that the curvature effect[6] can be used to interpret the pulse decay (including both prompt emission and X-ray flare) if the emission region moves with a constant Lorentz factor. One can measure the decay index ($\hat{\alpha}$) and the spectral index ($\hat{\beta}$) during the pulse decay, and they satisfy a simple relationship[7] ($\hat{\alpha} = 2 + \hat{\beta}$) in the Lab frame (Kumar & Panaitescu 2000). However, it is difficult to interpret the observed spectral lags of pulses with the curvature effect (Uhm & Zhang 2016b). Indeed, Uhm & Zhang (2016b) found that the dissipation of magnetic field energy in the shell via reconnection of magnetic field lines would result in a faster decrease than indicated by flux conservation if the magnetic field strength in the emitting region decreases with radius. This means that the emission region does not have a constant Lorentz factor but is accelerated, and the decay slope $\hat{\alpha}$ should be steeper than $2 + \hat{\beta}$ (Uhm & Zhang 2015). Afterwards, evidence for rapid bulk acceleration was discovered in observations of both X-ray flares (Jia et al. 2016; Uhm & Zhang 2016a) and GRB prompt emission (Uhm & Zhang 2016b; Li & Zhang 2021). In fact, Jia et al. (2016) found that a large fraction of X-ray flares in GRBs are inconsistent with the predicted relation from the curvature effect. Li & Zhang (2021) invoked the same method to analyze the prompt emission lightcurves of single-pulse GRBs, and suggested that the emission region of at least some GRBs is undergoing acceleration during the prompt emission phase.

A bright short-duration GRB 201221D, triggered the Swift Burst Alert Telescope at 23:06:34 UT on 2020 December 21 (BAT; Krimm et al. 2020) and located the source at R.A. = $11^{h}24^{m}12^{s}$ and decl. = $+42°08'39''$ (J2000). This GRB was also detected by the Fermi Gamma-ray Burst Monitor (GBM; Hamburg et al. 2020) and Konus-Wind (Frederiks et al. 2020). Based on spectroscopy of the optical counterpart, it was measured at a redshift of 1.046 (de Ugarte Postigo et al. 2020) which is larger than 95% of short-duration GRBs without extended emission (Dichiara et al. 2020). In order to test the physical origin of the high-$z$ short-duration GRB 201221D, we perform a comprehensive analysis of Fermi and Swift data on this burst as shown in Section 2. In Section 3, we compare some statistical relations of this burst with those of other long and short GRBs, and discuss its physical origin. In Section 4, we discuss the possible bulk acceleration in the prompt emission of this burst. A summary and conclusions are presented in Section 5. Throughout the paper, a concordance cosmology with parameters $H_0 = 71$ km s$^{-1}$ Mpc$^{-1}$, $\Omega_M = 0.30$, and $\Omega_\Lambda = 0.70$ is adopted.

## 2. Data Reduction and Analysis

### 2.1. Swift Data Reduction

GRB 201221D first triggered the Swift/BAT at 23:06:34 UT on 2020 December 21 (Page et al. 2020). The BAT data were processed using the HEASOFT package (v6.28). The light curves in different energy bands and spectra were extracted by running batbinevt (Sakamoto et al. 2008). The time bin size is fixed to 64 ms in this case due to the short duration, and the light curve shows a short-pulse with duration $T_{90} = 0.15 \pm 0.04$ s in the 15–350 keV (see Figure 1). The time-averaged spectrum from $T_0 - 0.06$ to $T_0 + 0.17$ s is best fit by a simple power-law model with spectral index $1.56 \pm 0.13$ due to the narrow energy band. Moreover, we do not find any signature of extended emission even up to 100 s following the burst. The X-ray Telescope (XRT) began observing the field at 87 s after the BAT trigger, but the source is too faint to be detected with photon counting (Evans et al. 2020).

### 2.2. Fermi Data Reduction

We downloaded the corresponding Time-Tagged-Event data from the public data site for Fermi/GBM.[8] For more details on the light curve and spectra data reduction procedure see the discussion in Zhang et al. (2016). The light curves of the n8 and b1 detectors are shown in Figure 1, and the background is modeled via applying the "baseline" method (Zhang et al. 2011) to a wide time interval before and after the signal and subtracting the GBM light curve. The lightcurves show a single-pulse emission with a duration of $T_{90} = 0.13 \pm 0.01$ in 50–300 keV. No significant signatures of precursor emission before the burst and extended emission (EE) after the burst are found in the GBM temporal analysis.

---

[6] The curvature effect is due to the observer receiving progressively delayed emission from higher latitudes (Liang et al. 2006; Zhang et al. 2007).
[7] Throughout the paper, the notation $f_\nu(t) \propto t^{-\hat{\alpha}} \nu^{-\hat{\beta}}$ is adopted.
[8] https://heasarc.gsfc.nasa.gov/FTP/fermi/data/gbm/daily/

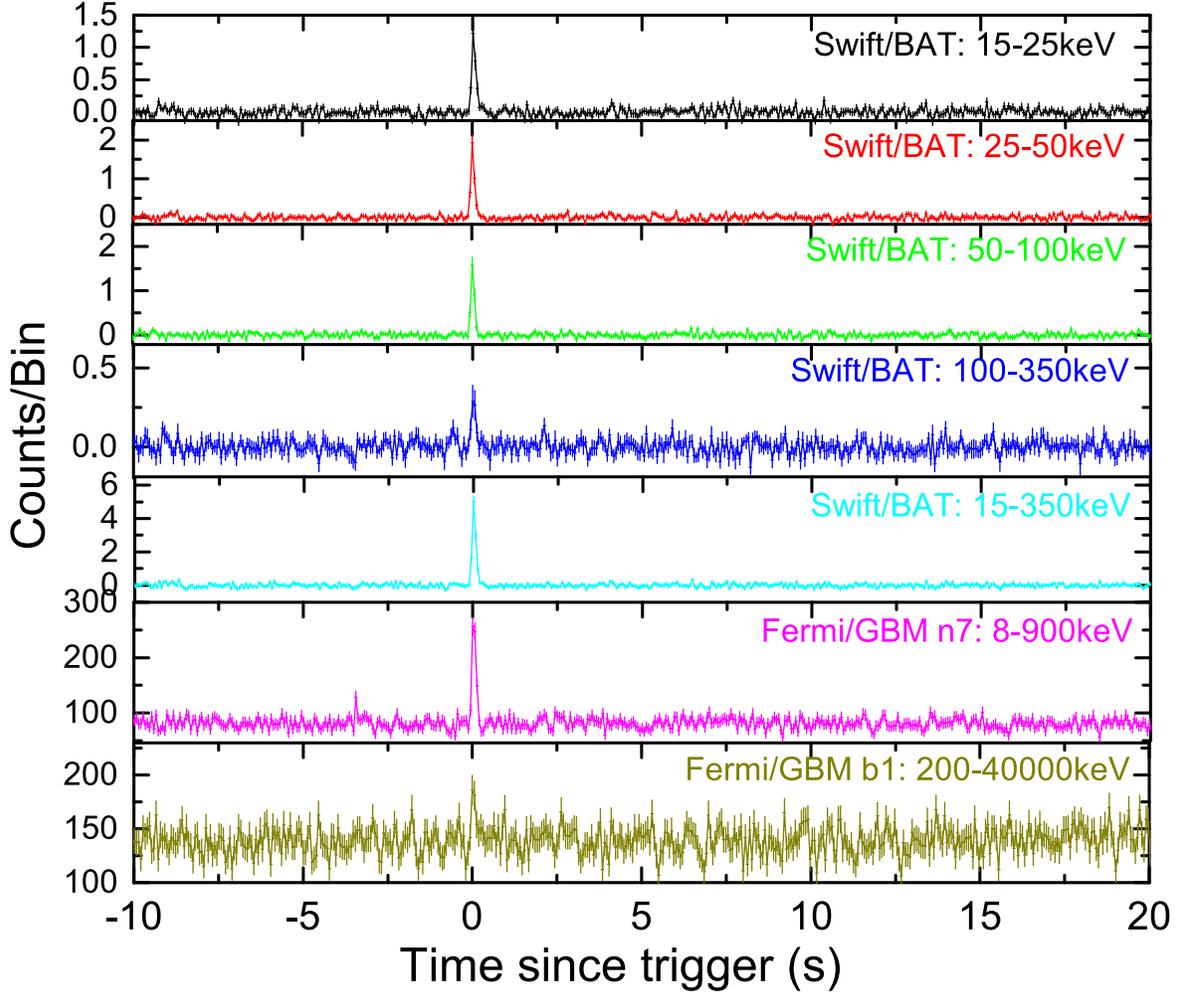

**Figure 1.** Swift/BAT and Fermi/GBM light curves of GRB 201221D in different energy bands with a 64 ms time bin.

We also extract both time-integrated and time-dependent spectral analyses of GRB 201221D between $T_0 - 0.03$ and $T_0 + 0.1$. This time interval is divided into four slices (see Table 1) based on brightness and the count statistical significance of the spectral fitting (Zhang et al. 2016, 2018a). The background spectra are extracted from the time intervals before and after the prompt emission phase and modeled with an empirical function (Zhang et al. 2011), and the spectral fitting is performed by using our automatic code "McSpecfit" in Zhang et al. (2018a). Several spectral models can be selected to test the spectral fitting of the burst, such as power-law (PL), cutoff power-law (CPL), Band function (Band), and Blackbody (BB). In order to test which model is the best fit of the data, we invoke the Bayesian Information Criteria (BIC)[9] to judge the best model among different models. The comparison of the goodness of the fits for different models is shown in Table 2. We find that the CPL model is the best one for adequately describing the observed data. The CPL model fit of the time-integrated spectrum is shown in Figure 2 for parameter constraints of the fit. It gives peak energy $E_p = 113^{+9}_{-7}$ keV, and a lower energy spectral index of $\Gamma = -0.26^{+0.21}_{-0.18}$. The best-fit parameters of the CPL fits are listed in Table 3. The CPL model is expressed as

$$N(E, t) = N_0(t) \cdot E^{-\Gamma} \exp\left(-\frac{E}{E_p}\right) \quad (1)$$

where $\Gamma$ and $N_0$ are the photon index and the CPL spectral fitting normalization, respectively. To extract the time-dependent spectrum, we use a similar method to the one mentioned above. We find that the CPL model is also the best fit, and the fitting results are shown in Table 3. One can see that the tracking spectral evolution is observed during the burst (see Figure 3).

**Table 1**
Properties of GRB 201221D

| | |
|---|---|
| $T_{90}$ (s) | $0.13 \pm 0.01$ |
| Redshift ($z$) | 1.046 |
| Spectral peak energy $E_p$ (keV) | $113^{+9}_{-7}$ |
| Total fluence (erg cm$^{-2}$) | $4.98^{+0.62}_{-0.53} \times 10^{-7}$ |
| Total isotropic energy $E_{\gamma,\mathrm{iso}}$ (erg) | $1.36^{+0.17}_{-0.14} \times 10^{51}$ |
| Luminosity $L_{\gamma,\mathrm{iso}}$ (erg s$^{-1}$) | $2.09^{+0.22}_{-0.26} \times 10^{52}$ |
| $f$-parameter | 2.13 |
| $\epsilon$-parameter | 0.034 |
| Event rate density (Gpc$^{-3}$ yr$^{-1}$) | $\sim 0.003$ |
| Offset $R_\mathrm{off}$ (kpc) | $4.50^{+3.64}_{-0.64}$ |
| Half light radius $R_{50}$ (kpc) | 4.4 |
| $R_\mathrm{off}/R_{50}$ | $1.01^{+1.84}_{-0.15}$ |
| Cumulative light fraction $F_\mathrm{light}$ | $0.38^{-0.37}_{+0.44}$ |
| Stellar mass $M_*$ ($M_\odot$) | $(3.9 \pm 3.1) \times 10^9$ |
| $\log P_\mathrm{II}/P_\mathrm{I}$ | $-5.53^{+1.14}_{-1.36}$ |

Table 2
BIC Values for Different Models We Adopted to Fit within Time-dependent Spectral Fitting

|  | BIC | | | | |
|---|---|---|---|---|---|
|  | (−0.03–0.01) s | (0.01–0.035) s | (0.035–0.0675) s | (0.0675–0.1) s | (−0.03–0.1) s |
| CPL | 180.34 | 154.91 | 190.35 | 185.03 | 218.30 |
| Band | 185.78 | 160.74 | 195.74 | 190.91 | 224.17 |
| PL | 245.46 | 166.28 | 263.77 | 199.82 | 440.44 |
| BB | 186.09 | 160.80 | 193.48 | 190.93 | 221.04 |

### 2.3. Host Galaxy and Other High-z Short GRBs

GRB 201221D was initially localized to R.A. = $11^h\ 24^m\ 14^s.19$, decl. = $+42^d\ 08'35''.5$ with $3''.9$ uncertainty by Swift/XRT (Evans et al. 2020). Later, with the $r'$ band afterglow from GTC/OSIRIS, it was localized to R.A. = $11^h\ 24^m\ 14^s.09$, decl. = $+42^d\ 08'40''.0$ with $1''$ uncertainty (Agüí Fernández et al. 2021). A faint galaxy around it was identified as the host galaxy. Kilpatrick et al. (2020) analyzed stacked images of the Pan-STARRS data release (Flewelling & Alatalo 2016), and found the galaxy to have a g-band magnitude $g = 23.2 \pm 0.2$ mag. This source was also observed by the Nordic Optical Telescope with $r = 23.1 \pm 0.3$ mag (Malesani & Knudstrup 2020) and the Lowell Discovery Telescope with $r = 23.9$ mag (Dichiara et al. 2020) which is consistent with emission identified in Kilpatrick et al. (2020). Recently, Agüí Fernández et al. (2021) observed the host and afterglow of GRB 201221D with GTC, and measured the redshift $z = 1.045 \pm 0.0008$ which is consistent with the GCN report. In addition, it is detected in the Dark Energy Spectroscopic Instrument Legacy Survey (DESI/LS), and listed in the LS DR8 catalog. The DESI/LS g-band image with the localization from Swift/XRT (white dashed circle), the optical afterglow (green circle and dot), and the host galaxy (red cross) are presented in Figure 4. We analyzed the images and the host galaxy properties as follows.

1. Offset and $R_{50}$: We use the position identified by SExtractor[10] to be the center of the host galaxy, R.A. = 11:24:14.04, decl. = +42:08:39.985. The offset of the r-band afterglow from the host center is $0.56^{+1.01}_{-0.08}$, which corresponds to $R_{off} = 4.50^{+3.64}_{-0.64}$ kpc at redshift 1.046. We used the half light radius of the host galaxy in Dark Energy Spectroscopic Instrument Legacy Survey (DESI/LS) catalog, which is $0''.55$. It corresponds to 4.4 kpc for the host galaxy redshift 1.046, and the normalized offset is $r_{off} = R_{off}/R_{50} = 1.01^{+1.84}_{-0.15}$.

2. $F_{light}$: We estimate the cumulative light fraction $F_{light}$, the fraction of the total brightness of the regions fainter than the GRB position to the total brightness of the host with the DESI image. Following Lyman et al. (2017), we use SExtractor (see footnote 10) to get the region of the host galaxy, and then sort the brightness of the pixels to estimate the fractional brightness of the regions fainter than the GRB 201221D region to the brightness of the host. The cumulative light fraction is estimated to be $F_{light} = 0.38^{+0.47}_{-0.44}$.

3. Stellar mass: We performed Spectral Energy Distribution (SED) fitting with the Code Investigating GALaxy Emission (CIGALE; Noll et al. 2009)[11] to estimate the host galaxy stellar mass. The DESI $g$, $r$, $z$ bands and Pan-STARRS $y$ band magnitude from Kilpatrick et al. (2020) are used, after the Galactic extinction correction (Schlafly & Finkbeiner 2011). The Chabrier model is used as the initial mass function, and the sfhdelayed model is used for star formation history with an initial SFR of 0.1 $M_\odot$ yr$^{-1}$. For the spectrum, we used *BC03* stellar population model (Bruzual & Charlot 2003) with the dustatt_calzleit dust attenuation model and the UV bump centroid to be 217.5 nm, as well as the Dale2014 dust emission model

---

[9] BIC is a criterion for model selection among a finite set of models, and it is defined as BIC = $\chi^2 + k \cdot \ln(n)$, where $k$ is the number of model parameters, and $n$ is the number of data points. The model with the lowest BIC is preferred.

[10] https://astromatic.net/software/sextractor/
[11] cigale.oamp.fr

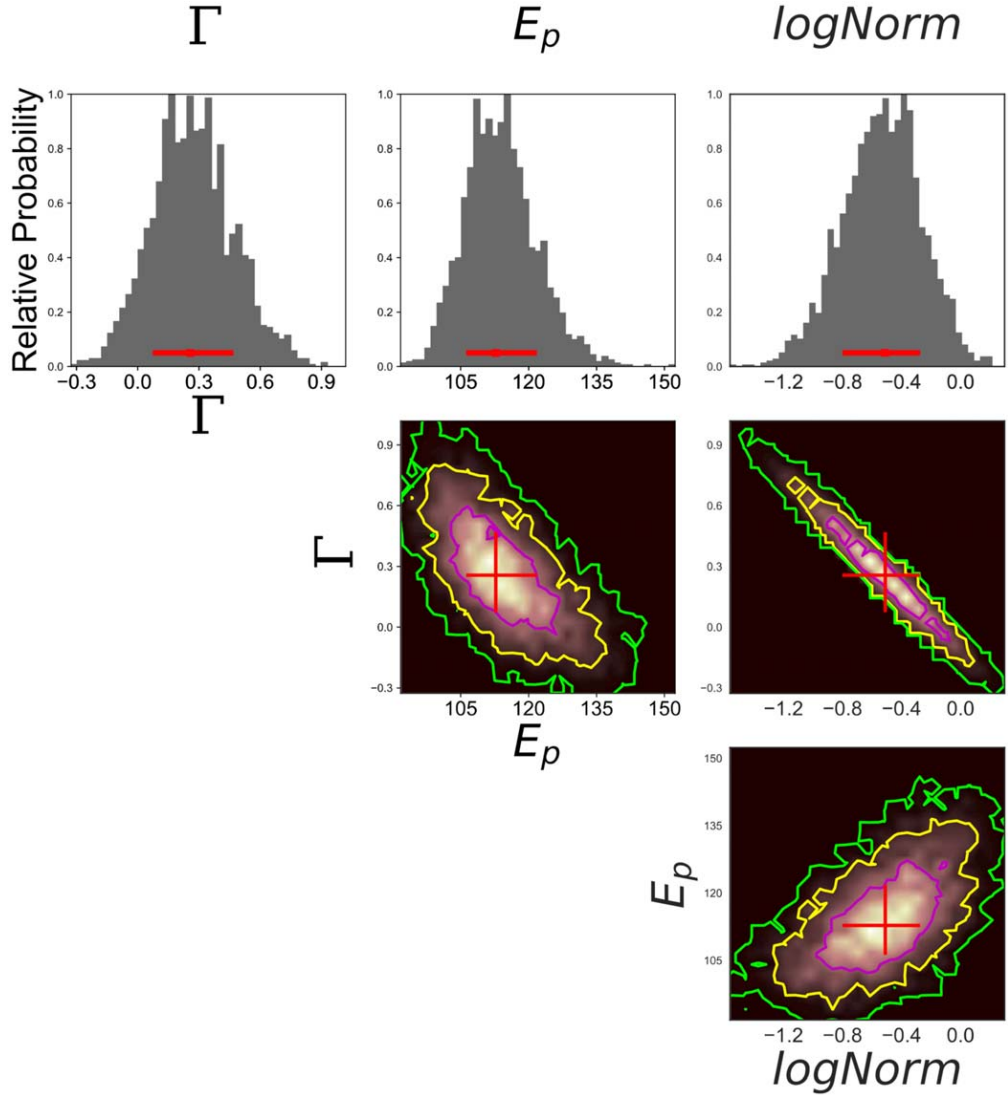

**Figure 2.** The parameter constraints of the spectral fit with the CPL model for GRB 201221D. Histograms and contours in the corner plots show the likelihood map of constrained parameters by using our McSpecFit package. Red crosses are the best-fitting values, and pink, yellow, and green circles are the 1$\sigma$, 2$\sigma$, and 3$\sigma$ uncertainties, respectively.

(Dale et al. 2014) and a various AGN fraction. The resulting stellar mass is $M_* = (3.9 \pm 3.1) \times 10^9 \, M_\odot$.

de Ugarte Postigo et al. (2020) obtained spectroscopy of the optical counterpart of GRB 201221D with the 10.4 m Telescope, and measured the redshift at $z = 1.046$ based on a prominent emission feature. The high-$z$ short GRBs play an important role in understanding the age of stellar progenitors, the cosmic chemical evolution, and formation channels of binary systems if we believe that short GRBs originate in the mergers of compact stars (Zheng & Ramirez-Ruiz 2007; Dominik et al. 2012; Anand et al. 2018). To date, more than 130 short GRBs have been detected by Swift/BAT, but less than 5% are found at $z > 1$ (Dichiara et al. 2021).

Dichiara et al. (2021) studied the properties of high-$z$ short GRBs with $z > 1$. They find that there are eight short GRBs with redshift $z > 1$, with five short GRBs having EE and three short GRBs (GRBs 090426, 111117A, and 121226A) without EE. However, the short-duration GRB 090426 with $z = 2.609$ seems to originate in a massive star core-collapse based on the properties of host galaxy and afterglow (Antonelli et al. 2009; Thöne et al. 2011; Xin et al. 2011).

### 2.4. Burst Energy

Based on the spectral analyses, one can estimate both event fluence and flux which are derived from the best model (CPL) within 1–$10^4$ keV during the time interval as

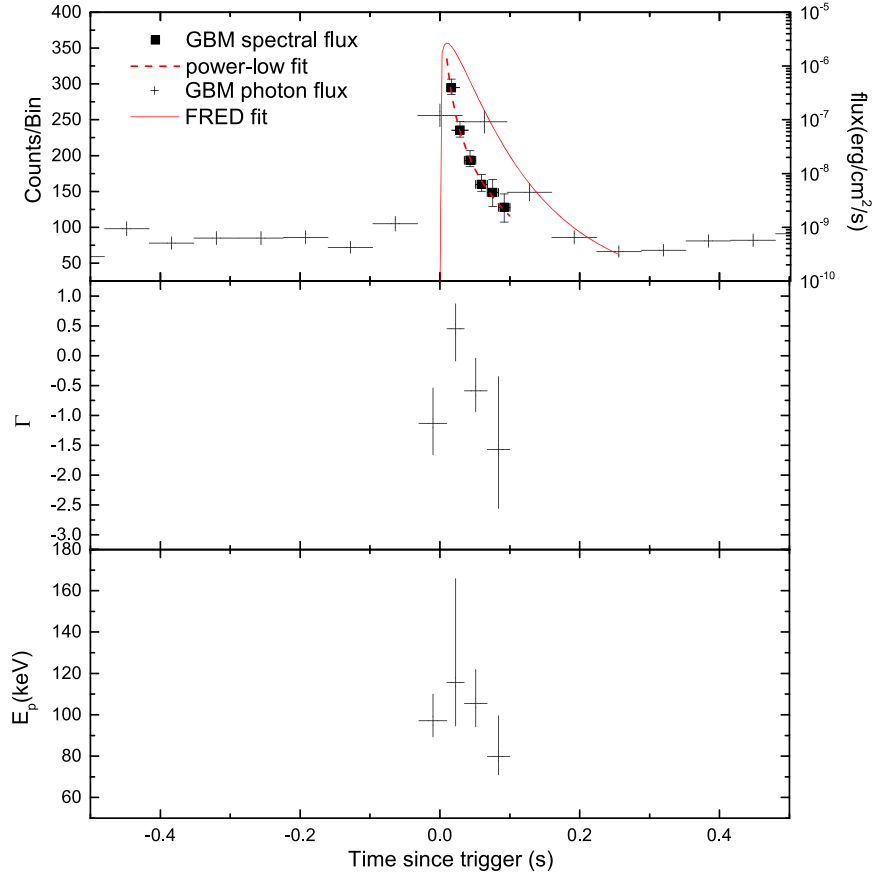

**Figure 3.** Zooming in to the GBM lightcurve of the pulse with photon and energy flux in time interval $[T_0 − 0.4, T_0 + 0.5]$, FRED fit (red solid line), and power-law fit (red dashed line), respectively (Top panel). The evolution of photon index (middle panel) and peak energy (bottom panel) with CPL model is presented.

**Table 3**
Time-dependent Spectral Fitting Results of GRB 201221D with CPL Model

| $t_1(s)$ | $t_2(s)$ | $\Gamma$ | $E_p$ (keV) | $\log A$ | BIC |
|---|---|---|---|---|---|
| −0.03 | 0.01 | $-1.13^{+0.59}_{-0.53}$ | $97.10^{+12.93}_{-7.85}$ | $-1.72^{+0.70}_{-0.83}$ | 180.34 |
| 0.01 | 0.035 | $0.45^{+0.42}_{-0.54}$ | $115.72^{+50.21}_{-21.19}$ | $0.50^{+0.71}_{-0.57}$ | 154.91 |
| 0.035 | 0.0675 | $-0.59^{+0.55}_{-0.35}$ | $105.49^{+16.45}_{-11.37}$ | $-0.89^{+0.46}_{-0.77}$ | 190.35 |
| 0.0675 | 0.1 | $-1.57^{+1.22}_{-0.99}$ | $79.90^{+19.59}_{-9.00}$ | $-2.41^{+1.25}_{-1.73}$ | 185.03 |
| −0.03 | 0.1 | $-0.26^{+0.21}_{-0.18}$ | $112.75^{+9.09}_{-6.51}$ | $-0.51^{+0.24}_{-0.29}$ | 218.30 |

$4.98^{+0.62}_{-0.53} \times 10^{-7}$ erg cm$^{-2}$ and $3.83^{+0.47}_{-0.4} \times 10^{-6}$ erg cm$^{-2}$ s$^{-1}$ respectively. By adopting basic cosmological parameters at redshift $z = 1.046$, the corresponding isotropic energy and peak luminosity are estimated as $E_{\gamma,\mathrm{iso}} = 1.36^{+0.17}_{-0.14} \times 10^{51}$ erg and $L_{\gamma,\mathrm{iso}} = 2.09^{+0.22}_{-0.26} \times 10^{52}$ erg s$^{-1}$, respectively. The above values are summarized in Table 1.

## 3. What is the Physical Origin of GRB 201221D?

The classification of GRBs remains an open question (Zhang 2011). Our purpose is to investigate the physical origin of GRB 201221D. In this section, we will discuss the origin of GRB 201221D by comparing the properties of GRB 201221D with those of other long- and short-duration GRBs, e.g., the Amati relation (Amati et al. 2002), luminosity function, properties of host galaxy, "tip of the iceberg" effect (Lü et al. 2014), and $\varepsilon$-classification method (Lü et al. 2010).

### 3.1. Comparisons of the Empirical Relationships of GRB 201221D with other Type I/II GRBs

#### 3.1.1. $E_p$–$E_{\gamma,iso}$ Relation

Amati et al. (2002) discovered that higher energy Type II GRBs have a harder spectrum than that of lower energy Type II GRBs, and the relationship is usually expressed in terms of $E_p \propto E_{\gamma,\mathrm{iso}}^{1/2}$. The Type I GRBs also follow the same trend between $E_p$ and $E_{\gamma,\mathrm{iso}}$ but form distinct tracks (Zhang et al. 2009). The $E_p$ of GRB 201221D is about 113 keV which is less than 90% of the Type I GRBs observed by Fermi/GBM (see Figure 5; Lu et al. 2017), but mixed in with most Type II GRBs (von Kienlin et al. 2020). In Figure 5, we also re-plot $E_p$–$E_{\gamma,\mathrm{iso}}$ for GRB 201221D to compare with other Type II and Type I GRBs in the rest frame. We find that GRB 201221D is located at the outlier of 3σ uncertainty of fits for both Type I and Type II GRBs, but seems to be closer to the distribution of Type I GRBs in comparison to Type II GRBs. It is difficult to judge the progenitor based only on this empirical correlation.

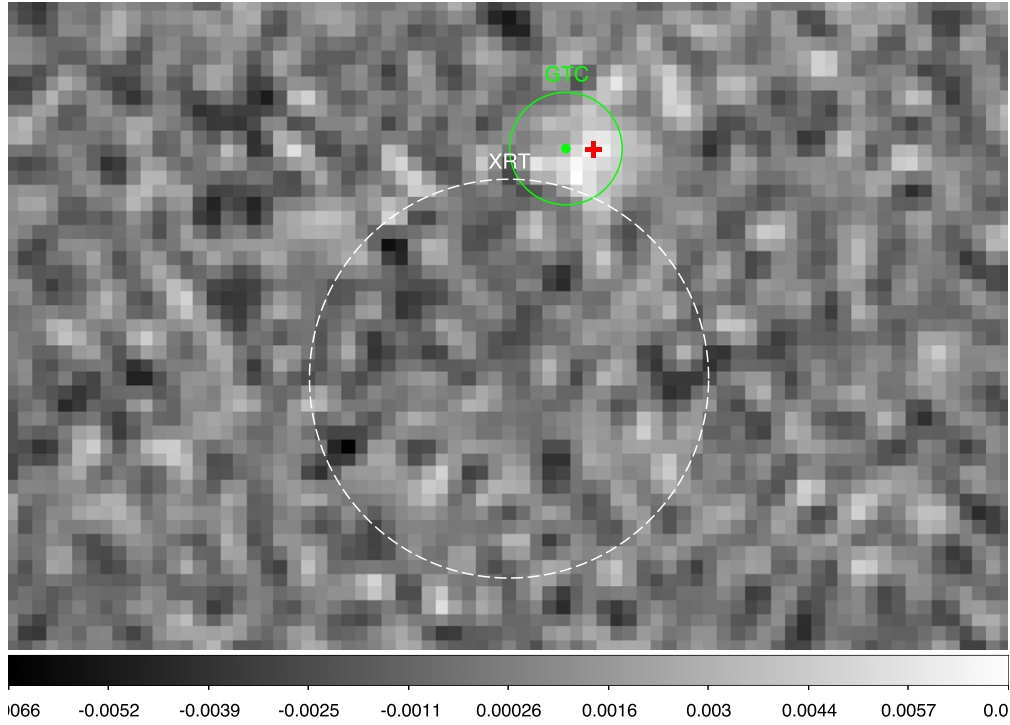

**Figure 4.** The DESI/LS *g*-band image with the localization of Swift/XRT (white dashed circle), optical afterglow (green circle and dot) and the host galaxy (red cross) of GRB 201221D.

### 3.1.2. Local Event Rate Density

The event rate density describes how many events happen per volume per unit of time. The observed event rate density is redshift-dependent, luminosity-dependent, and beaming-dependent. Most short-duration GRBs are believed to originate in the merger of compact stars, and the observed local event rate density has a large uncertainty. Sun et al. (2015) estimated the local event rate density of short GRBs with ∼(0.5–3) $Gpc^{-3}\,yr^{-1}$ above $10^{50}\,erg\,s^{-1}$. We estimate the local event rate density of the source from $\rho_0 = 1/(V_{max}T)$, where $V_{max}$ is the maximum volume from which the source can be detected weighted by the redshift evolution (see Equation (3) in Sun et al. 2015) and $T$ is the total exposure time of the telescope or survey. Assuming a flux limit of $F_{th} = 10^{-7}\,erg\,cm^{-2}\,s^{-1}$, and total operation time of 16 yr for Swift, the local event rate density of GRB 201221D is $\sim 3.0^{+6.9}_{-2.5} \times 10^{-3}\,Gpc^{-3}\,yr^{-1}$ for the peak bolometric luminosity of $\sim 2.09 \times 10^{52}\,erg\,s^{-1}$ in $1-10^4$ keV. The 1σ errors are derived from Gehrels (1986) based on one detection. This value is lower than that of other Type II GRBs within the same luminosity range. A comparison of the event rate density of GRB 201221D with other Type I/II GRBs is shown in Figure 6. Moreover, we do not consider the beaming factor effect in our calculations due to uncertainty of jet opening-angle in this case.[12]

### 3.1.3. Luminosity Function

The luminosity function of high-luminosity (HL) long-duration GRBs can be characterized as a broken power law function via a large enough sample of redshift measurements, but the luminosity function of low-luminosity (LL) long-duration GRBs is not well constrained with a small sample size due to the detectors' sensitivity limit (Liang et al. 2007; Virgili et al. 2009; Sun et al. 2015). In comparison with long-duration GRBs, the luminosity function of short-duration GRBs is less well constrained with a small fraction of the data. Sun et al. (2015) found that the luminosity function of short-duration GRBs can be roughly fitted with a simple power law in the luminosity range of $[7 \times 10^{49}, 10^{50}]\,erg\,s^{-1}$ by assuming that all short-duration GRBs have a compact star merger origin. GRB 201221D broadly follow the distributions of both the long and short GRB populations, as shown in the right panel of Figure 6. Moreover, Paul (2018) analyzes the luminosity function with a large sample of short-duration GRBs observed by CGRO/BATSE, Swift/BAT, and Fermi/GBM. They found that the luminosity function can be described with the exponential cutoff power law and broken power law models. However, the redshifts of most short GRBs in their samples are not measured, and they adopt pseudo-redshifts which are derived by the empirical relationship. Therefore, we still used the results from Sun et al. (2015).

---

[12] By considering the effect of beaming factor, one can roughly estimate local event rate densities which are 25 and 500 times greater than those ignoring the beaming factor, respectively when we adopt the typical beaming factors ∼0.04 and ∼0.002 for short- and long-duration GRBs cases (Frail et al. 2001; Fong et al. 2015). If this is the case, the local event rate density is nearly consistent with other Type I and II GRBs within the same luminosity range.

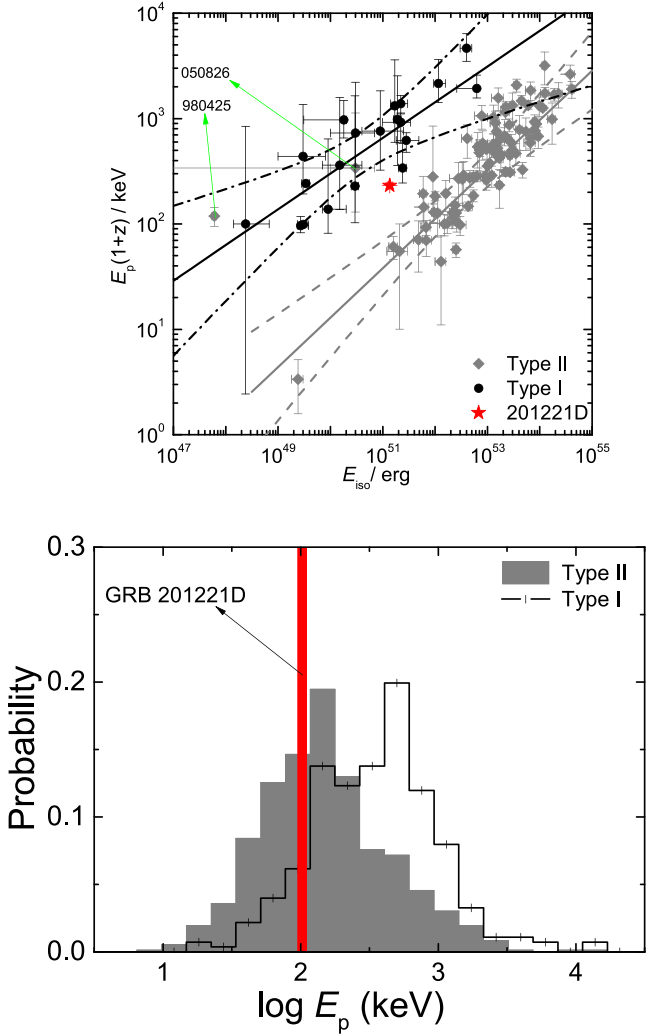

**Figure 5.** Top panel: $E_p$ and $E_{iso}$ correlation diagram. Black points and gray diamonds correspond to Type I and Type II GRBs, respectively. The red star is GRB 201221D, and other data are taken from Zhang et al. (2009). The best-fit $E_p$–$E_{iso}$ correlations for both Type II (gray diamonds) and Type I (black points) GRBs are plotted (solid lines) with the $3\sigma$ boundary (dashed line) marked. Bottom panel: $E_p$ distribution of GRB 201221D and other Type I/II GRBs observed by Fermi/GBM. The data of $E_p$ values of other Type I/II GRBs are taken from Lu et al. (2017) and von Kienlin et al. (2020), respectively.

### 3.1.4. ε-parameter

Lü et al. (2010) proposed a new phenomenological classification method for GRBs by introducing a new parameter $\varepsilon$, which is defined as $\varepsilon = E_{\gamma,\mathrm{iso},52}/E_{p,z,2}^{5/3}$ for GRBs with $z$ measurements by considering both the burst energy and the spectral properties in the burst rest frame. Here, $E_{\gamma,\mathrm{iso},52}$ is the isotropic gamma-ray energy in units of $10^{52}$ erg) and $E_{p,z,2}$ is the cosmic rest-frame spectral peak energy in units of 100 keV. They found that the $\varepsilon$ parameter shows a clear bimodal distribution with a separation at $\varepsilon \sim 0.03$ by invoking the current complete sample of GRBs with redshift and $E_p$ measurements. This method can separate very well the observed GRBs as high-$\varepsilon$ and low-$\varepsilon$ regions that correspond to massive star core-collapse (Type II) and merger of two compact stars (Type I) origin, respectively. As a caution, an imperfection of this method is that it is not good enough for those cases of short GRBs with extended emission and low-luminosity GRBs (see the green triangles in Figure 7). By adopting this method for GRB 201221D, one can calculate $\varepsilon \sim 0.034$ which lies precisely on the separation line (see Figure 7). It remains to carry a confused information to distinguish the physical origin of GRB 201221D.

### 3.1.5. Amplitude Parameter

Lü et al. (2014) suggested that the amplitude of an observed lightcurve may be taken into account as a third dimension in classifying GRBs. They define $f$ parameter as the ratio between the peak flux and the background flux of a GRB. This parameter reflects the apparent brightness of a GRB, and tells us how bright it is above the background. This parameter alone does not help us much to understand the origin of a GRB because the distributions of $f$ values for long- and short-duration GRBs are similar. They also define another parameter $f_{\mathrm{eff}}$ as the ratio between pseudo-peak flux and average background flux. The pseudo-peak flux is defined as follows: for each long-duration GRB, one can simulate a pseudo GRB by scaling down the flux globally, until the signal above the background has a duration shorter than 2 s. We called it a "pseudo short-duration GRB" from a long-duration GRB. The amplitude parameter of the pseudo GRB is defined as $f_{\mathrm{eff}}$ of the original long-duration GRB. Its physical meaning is that a long-duration GRB is confused as a short-duration GRB because most of its emission is buried beneath the background, so we called this effect the "tip-of-iceberg" effect. One can say that the amplitude of a "disguised short-duration" GRB may be due to the "tip-of-iceberg" effect. Nonetheless, contamination from long-duration GRBs indeed happens when the observed $f$ value of a short-duration GRB is small, and one can calculate that the contamination probability rapidly increases with decreasing $f$ as $P(<f) \sim 0.78^{+0.71}_{-0.4} f^{-4.33 \pm 1.84}$ when we set up GRB 090426 as the standard short-duration GRB from a Type II origin.

In order to test whether the short-duration GRB 201221D is a "tip-of-iceberg" of long-duration GRB, we performed the same analysis and obtained its amplitude parameter as $f = f_{\rm eff} \sim 2.13$. Figure 7 shows $T_{90}$ as a function as $f$ and $f_{\rm eff}$ for both Type II and Type I GRBs, as well as GRB 201221D. The $f_{\rm eff}$ value of GRB 201221D is comparable with the average values of other Type I GRBs, and is larger than that of most Type II GRBs. Moreover, we also calculated the probability of a disguised short-duration GRB according to the $p$–$f$ relation derived by Lü et al. (2014). We find that the probability ($p$) for it to be a disguised short-duration GRB is $p \sim 0.03^{+0.03}_{-0.02}$, meaning that GRB 201221D is more likely an intrinsic Type I GRB with a compact star merger origin.

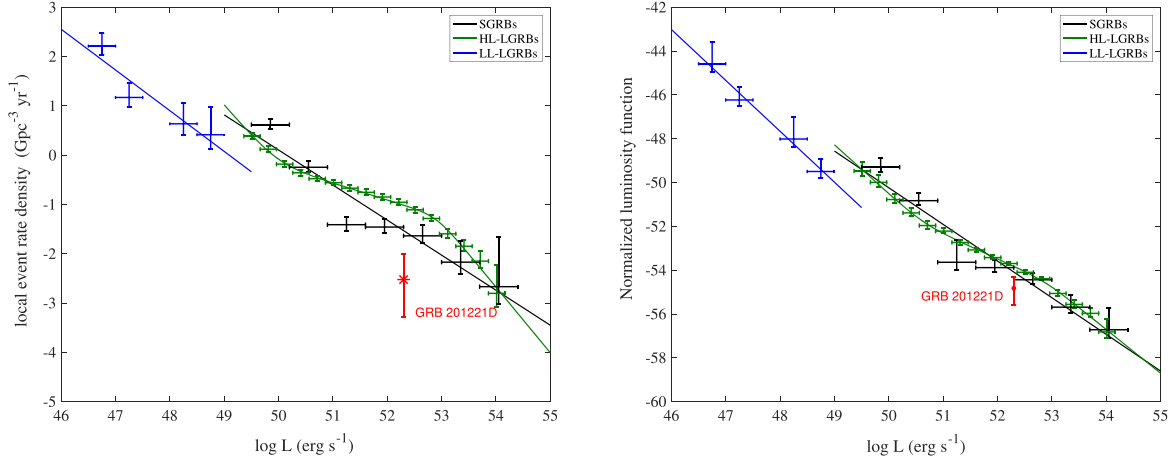

**Figure 6.** Local event rate density (top) and normalized luminosity function (bottom) distribution for GRB 201221D (red), LL-LGRBs (blue), HL-LGRBs (olive), and short-duration GRBs (black) inferred Sun et al. (2015).

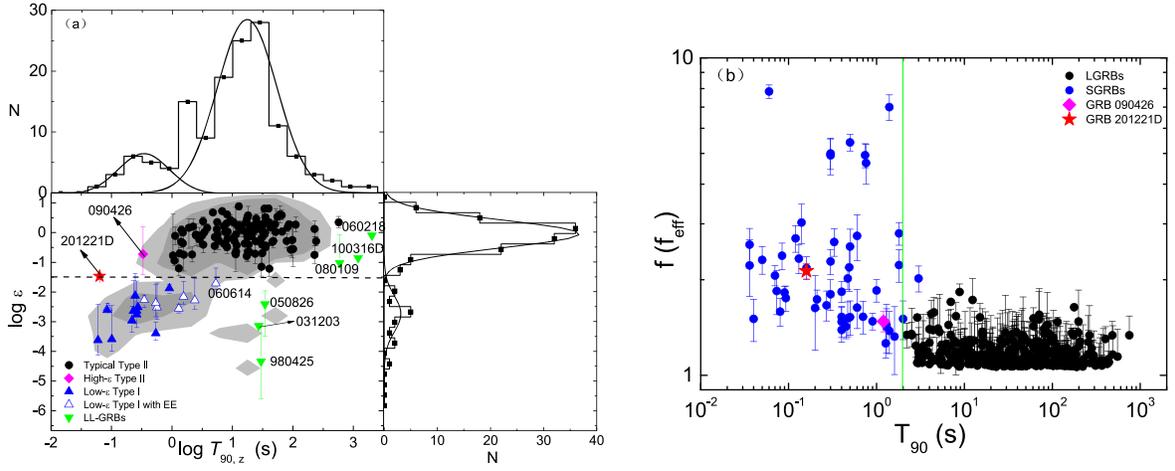

**Figure 7.** (a) 1D and 2D distributions of GRBs samples in $T_{90}$–$\varepsilon$ space. The dotted line is $\varepsilon = 0.03$. (b) $f(f_{\rm eff})$–$T_{90}$ for type I and Type II GRBs which are from Lü et al. (2014). The vertical solid line is $T_{90} = 2$ s. Blue (black) solid circles represent the Type I(II) GRB candidates, and green triangles denote the nearby low-luminosity long GRBs. The red star is GRB 201221D.

### 3.1.6. Host Galaxy Properties—the Probability to Be a Type I GRB—log Odds

Statistically speaking, the properties of the host galaxy for long and short GRBs are different (Bloom et al. 1999; Li et al. 2016), and the host galaxy and position information of the GRB inside the galaxy often gives clues regarding the progenitor of a GRB (Li & Zhang 2020). In order to test the origin of GRB 201221D, we compare the host galaxy properties of it with both type II (black) and type I (blue) GRBs (see Figure 8). Furthermore, we use the Naive Bayes method suggested by Li & Zhang (2020) to examine the similarity quantitatively. Naive Bayes method is a Bayesian theory based multivariate classifier, assuming parameters are independent. It estimates the likelihood of GRB 201221D to be either a type I or II GRB, $P_{\rm I}$ and $P_{\rm II}$, as the products of the likelihood to have each parameter. Following Li & Zhang (2020), the prompt emission properties include duration $T_{90}$, isotropic energy $E_{\rm iso}$, the lower energy index $\Gamma$ and peak energy $E_{\rm p}$ of the cutoff power-law for spectral modeling, and the effective amplitude parameter $f_{\rm eff}$ (Lü et al. 2014). The host galaxy properties include the stellar mass $M_*$, half light radius $R_{50}$ of the host galaxy, the normalized offset from the center of the host to the GRB $r_{\rm off} = R_{\rm off}/R_{50}$, and the cumulative light fraction $F_{\rm light}$. It turns out that the probability of GRB 201221D being a type II or type I GRB is $P_{\rm II} = 2.9 \times 10^{-6}$ and $P_{\rm I} = (1-2.9) \times 10^{-6}$, respectively. In addition, Li & Zhang (2020) suggests calculating the ratio between $P_{\rm I}$ and $P_{\rm II}$ to quantitatively examine the preference of one GRB to be type I and type II. The logarithmic probability ratios $\log {\rm Odds} = \log P_{\rm II}/P_{\rm I} = -5.53^{+1.14}_{-1.36}$, which prefers type I significantly. If we consider the prompt emission properties and host galaxy properties separately, $\log {\rm Odds}_{\rm prompt} = -5.55^{+0.25}_{-0.15}$ and $\log {\rm Odds}_{\rm host} = 1.08^{+1.09}_{-1.40}$. This indicates that the prompt emission properties significantly prefer type I.

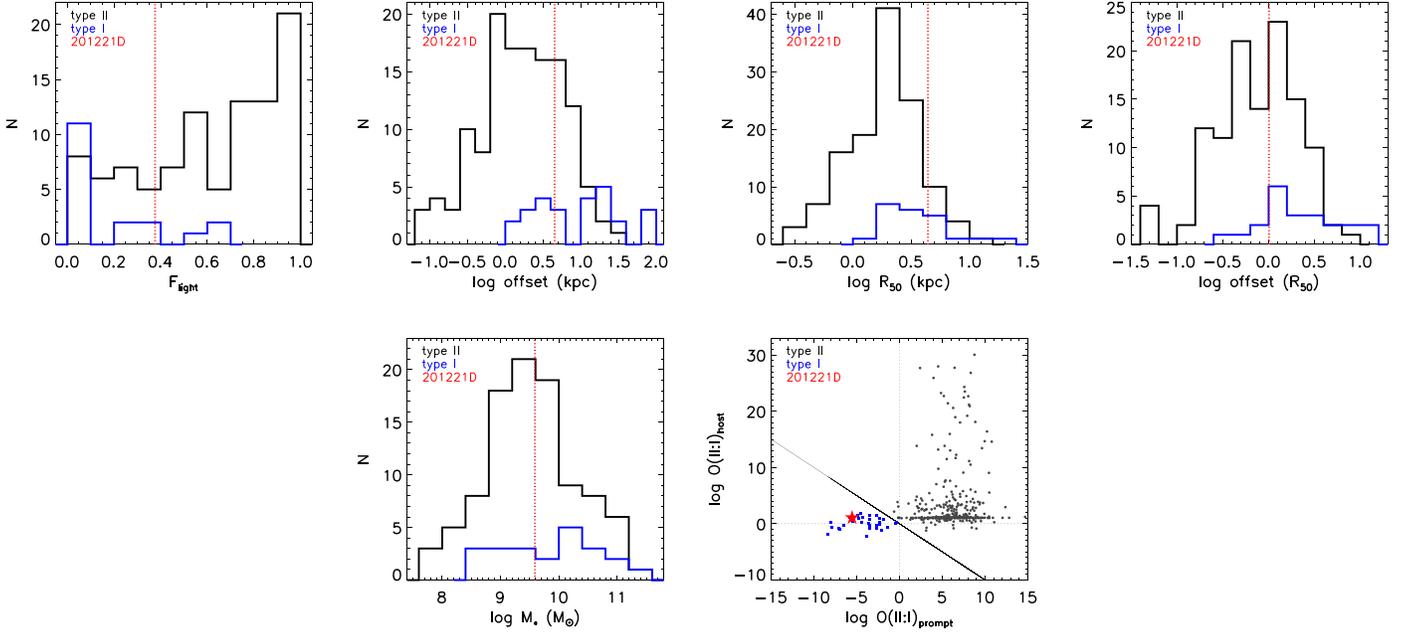

**Figure 8.** Host galaxy properties of GRB 201221D. Distributions of cumulative light fraction $F_{\rm light}$, physical offset, normalized offset, stellar mass, and posterior odds. Black and blue histograms are for Type I and Type II GRBs, respectively. The vertical dashed red lines represent the location of GRB 201221D.

### 3.2. Merger of Compact Star Scenario

A leading progenitor for producing short-duration GRBs is binary neutron star mergers which were confirmed via the detection of the associated of gravitational-wave (GW) 170817 and GRB 170817A (Abbott et al. 2017; Zhang et al. 2018b). Neutron stars likely receive a kick at birth from SNe explosions, and mergers of binary neutron stars are usually found to have a large offset from the galaxy center or even be outside of the host galaxy (Bloom et al. 1999; Li & Zhang 2020). As a caution, Mandhai et al. (2021) point out that the observational offset is the minimum value due to projection effects, namely the real-offset is much larger. Most observations (amplitude parameter, properties of host galaxy, local event rate) support a compact star merger origin.

Moreover, black hole and white dwarf merger systems were also proposed as possible progenitors of short-duration GRBs (Fryer et al. 1999), but these systems may not be favorable for launching a GRB jet with a small fraction of accreted mass (Narayan et al. 2001). MacFadyen et al. (2005) proposed that accretion-induced collapse of a neutron star can power a GRB, but this model can produce GRBs in non-star-forming galaxies with a small offset of GRB location. It is also inconsistent with the observations of the host galaxy of GRB 201221D. White dwarf-white dwarf mergers producing a short GRB were discussed by Lyutikov & Toonen (2017), but this model predicts the powering of long-lasting EE after the prompt emission, which is disfavored for GRB 201221D because there is no observed EE.

### 3.3. Massive Star Core-collapse Scenario

Alternatively, the short-duration GRB with high-redshift measurements may originate in the collapse of a massive star rather than a compact object, such as GRB 090426 (Levesque et al. 2010; Xin et al. 2011). Recently, Zhang et al. (2021) discovered that the short-duration GRB 200826A is more likely originates in a massive star core collapse. Inspired by this case, we are attempting to explain the observational properties of GRB 201221D by supposing that it has originated from collapse of a massive star. If GRB 201221D comes from a massive star collapse, the materials of the envelope fall to feed the central engine, and launch a jet with GRB emission when the massive star's core suddenly loses its pressure. If this is the case, the timescale of a powered relativistic jet is roughly equal to the freefall timescale of the star (Zhang 2018), it reads as $t_{\rm ff} \sim \left(\frac{3\pi}{32G\bar{\rho}}\right)^{1/2} \sim 210\,{\rm s}\,\left(\frac{\bar{\rho}}{100\,{\rm g\,cm^{-3}}}\right)^{-1/2}$, where $\bar{\rho}$ is the average density of the accreted materials (Zhang et al. 2021). By adopting a $\sim 0.1$ s duration of GRB 201221D, a lower limit on the density is found to be $4.4 \times 10^8\,{\rm g\,cm^{-3}}$, which is much higher than that of observations of massive star evolution, which seems to be inconsistent with the scenario of massive star collapse.

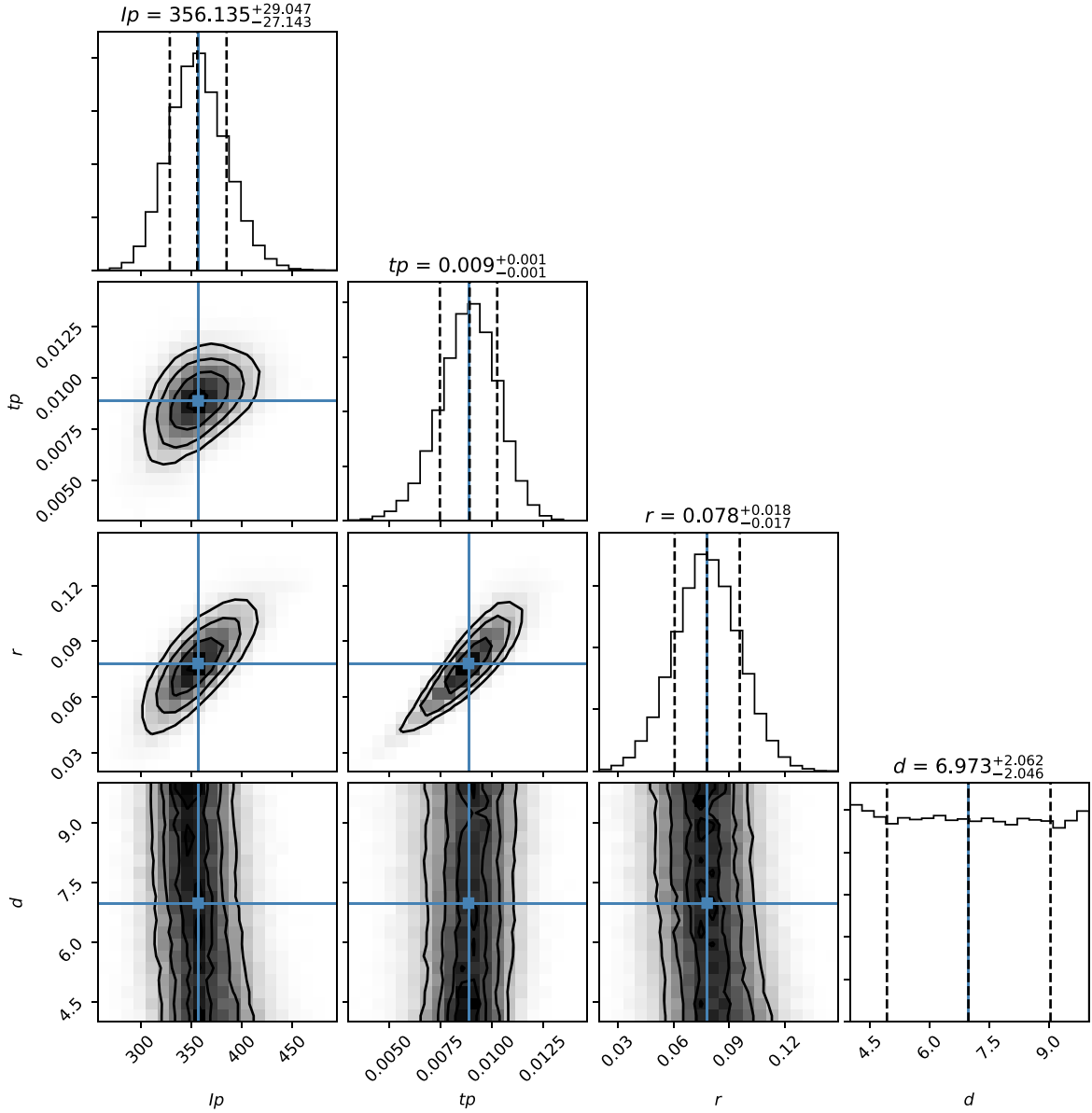

**Figure 9.** One- and two-dimensional projections of the posterior distributions of the Equation (2) free parameters by MCMC fit with pulse. Vertical dashed lines mark the the median and $1\sigma$ range. The contours are drawn at 68%, 95%, and 99% credible levels.

Such a puzzle might be solved in three ways if we believe that GRB 201221D originated from massive star core collapse. One is that GRB 201221D is produced by the collapse of a supramassive NS (or magnetar) into a black hole (Rezzolla & Kumar 2015), and the supramassive NS is the initial remnant of massive star collapse due to the stiff equation of state of NS, but does not power the GRB during this process. If this is the case, it should leave clues in the X-ray band when it loses rotational energy via dipole radiation. Unfortunately, there is not enough X-ray observational data after the prompt emission of GRB 201221D. The second possibility is that the total central engine timescale ($t_{\rm eng}$) is much longer than that of what we observed, but the majority of its time ($t_{\rm jet}$) is used in breaking the envelope out of the stellar surface, so the observed duration time of GRB ($t_{\rm GRB} = t_{\rm eng} - t_{\rm jet}$) could be as short as 1 s (Bromberg et al. 2012). However, this timescale is still one order of magnitude longer than the observed duration of GRB 201221D. The third possibility is that GRB 201221D is not intrinsically short but is actually long duration due to the tip of the iceberg effect (Lü et al. 2014). Within this scenario, the observed fluence is underestimated, which means that the intrinsic isotropic energy $E_{\gamma,\rm iso}$ should be larger than the current value what we calculated in GRB 201221D. If this is the case, the location of GRB 201221D in both $E_{\rm p}$–$E_{\gamma,\rm iso}$ and $\varepsilon$-parameter diagrams will move into the long GRB population. In that case, it is natural to explain the observed empirical correlations. However, the $f$-parameter of GRB 201221D should be consistent with that of other Type II GRBs, but the observed $f$-parameter of GRB 201221D does not support this hypothesis.

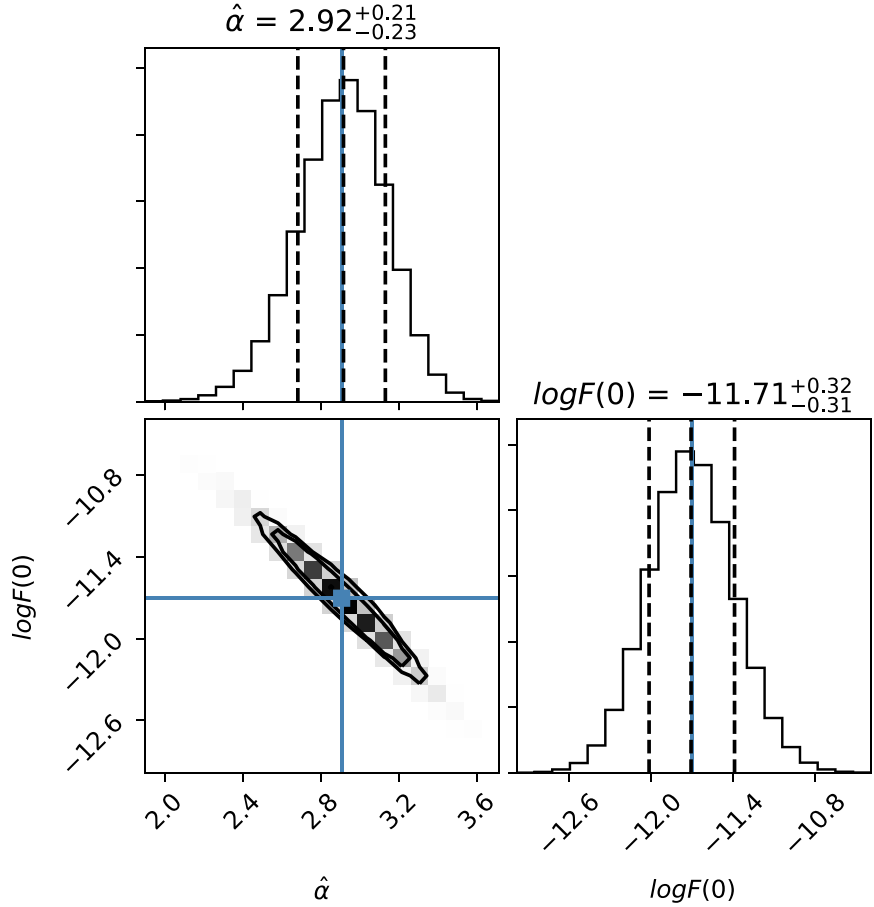

**Figure 10.** Similar with Figure 9, but adopting a power-law model to fit the decay segment of pulse.

## 4. Possible Bulk Acceleration in Prompt Emission?

Uhm & Zhang (2016b) proposed that the relationship between $\hat{\alpha}$ and $2 + \hat{\beta}$ is not consistent with the predicted of curvature effect when the emission region itself is undergoing acceleration or deceleration in the prompt emission phase. In this section, we describe how to test whether the prompt emission of GRB 201221D is undergoing acceleration or deceleration to diagnose the composition of the jet.

As the light curve of GRB 201221D shows a single-peaked structure, we adopt a fast-rising and exponential decay (FRED) empirical function to fit in order to describe the shape of a pulse (Kocevski et al. 2003). It reads as,

$$I(t) = I_p \left( \frac{t + t_0}{t_p + t_0} \right)^r \left[ \frac{d}{r+d} + \frac{r}{r+d} \left( \frac{t + t_0}{t_p + t_0} \right)^{r+1} \right]^{-\frac{r+d}{r+1}} \quad (2)$$

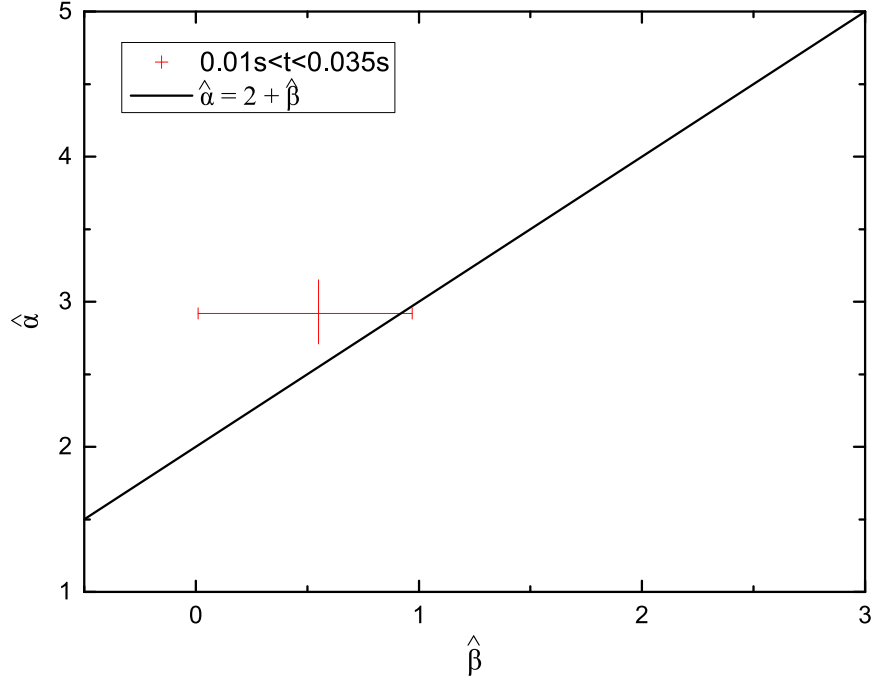

**Figure 11.** $\hat{\alpha}$–$\hat{\beta}$ relation of the curvature effect during the initial decay phase of prompt emission. The solid line is $\hat{\alpha} = 2 + \hat{\beta}$.

where $I_p$ is the intensity of amplitude, $r$ and $d$ are the rise and decay timescale parameters, and $t_0$ and $t_p$ are the zero time and peak flux of the pulse, respectively. Here, one always sets $t_0 = 0.0$ s due to the short-duration of the pulse itself[13] (see Figure 3). We invoke a Markov Chain Monte Carlo (MCMC) method with the python package emcee[14] to fit the lightcurve, and the fitting results are presented in Figure 9. One has the peak time $t_p = 0.009 \pm 0.001$ s.

In order to discover the relationship between $\hat{\alpha}$ and $\hat{\beta}$ during the decay phase of prompt emission, one needs to measure the decay slope $\hat{\alpha}$ and $\hat{\beta}$ in energy flux units, which is dependent on the evolved spectral index. Based on the time-dependent spectrum with CPL model fit, one has $\hat{\beta} = \Gamma + 1$. By adopting Equation (1), the photon flux can be converted to energy flux within the three time slices after the peak,[15] which reads as,

$$F(t) = N_0(t) \cdot E^{-\hat{\beta}} \exp\left(-\frac{E}{E_p}\right). \quad (3)$$

---

[13] This hypothesis is different from Jia et al. (2016) for X-ray flares. They adopted the first rising data point of each flare as the $t_0$ in X-ray flares, and extrapolating the rising light curve until it is three orders of magnitude lower than the peak flux density.
[14] https://emcee.readthedocs.io/en/stable/
[15] The larger uncertainty of the power-law fit with three flux points is not convincing. Here, we divided each time-bin which we used to analyze the time-resolved spectrum into two time slices, so that the three time-resolved spectral analysis bins correspond to six flux points.

Then, a power-law model, $F(t) = F(0)(t + t_0)^{-\hat{\alpha}}$, is invoked to fit the decay phase of prompt emission by using an MCMC fit with $\hat{\alpha} = 2.92^{+0.21}_{-0.23}$. The fitting results are shown in Figures 3 and 10. Three time intervals of prompt emission after the peak are adopted into the spectral analysis (in Table 2), and one may obtain $\hat{\beta}$ values due to the evolution of spectrum.

Figure 11 shows the correlation between $\hat{\alpha}$ and $\hat{\beta}$ during the decay phase of prompt emission. We find that $\hat{\alpha}$ is always larger than that of $2 + \hat{\beta}$. This result is not consistent with the predictions of the curvature effect, but strongly suggests that the emission region is undergoing bulk acceleration during the prompt emission phase. Based on the relationship between $L_{\gamma,\mathrm{iso}}$, $E_{p,z}$, and $\Gamma_{\mathrm{jet}}$ discovered by Liang et al. (2015), one can roughly estimate the $\Gamma_{\mathrm{jet}} \sim 450$ of GRB 201221D by adopting the calculated values of $L_{\gamma,\mathrm{iso}}$ and $E_{p,z}$. Moreover, we also calculate the distance of this region from the central engine, $R_{\mathrm{GRB}} = \frac{\Gamma_{\mathrm{jet}}^2 ct}{1+z} \sim 10^{16}$ cm with Lorentz factor $\Gamma_{\mathrm{jet}} \sim 450$ and typical $t \sim 0.1$ s. The large distance is not consistent with the predictions of the photosphere and internal shock models with thermally driven bulk acceleration (Kumar & Zhang 2015), but favors the Internal Collision-induced Magnetic Reconnection and Turbulence (ICMART; Zhang & Yan 2011) model. Within this scenario, the jet is Poynting-flux-dominated and undissipated until reaching a large enough distance (e.g., $\sim 10^{15}$ cm), and the dissipated energy is used to radiate $\gamma$-ray (prompt emission) and accelerate the ejecta. This conclusion is also consistent with results of GRB prompt emission in Li & Zhang (2021).

## 5. Conclusions

GRB 201221D is a short-duration burst with a $T_{90} \sim 0.1$ s in 50−300 keV at redshift $z = 1.046$, observed by both Swift and Fermi. We do not find any significant signatures of precursor emission before the burst and extended emission after the burst in both the Swift/BAT and Fermi/GBM temporal analyses. By extracting spectral analyses of GRB 201221D, we find that a cutoff power-law model can adequately fit the spectrum when time-integrated with a soft $E_p = 113^{+9}_{-7}$ keV. The isotropic energy ($E_{\gamma,\text{iso}}$) and peak luminosity ($L_{\gamma,\text{iso}}$) are estimated as $E_{\gamma,\text{iso}} = 1.36^{+0.17}_{-0.14} \times 10^{51}$ erg and $L_{\gamma,\text{iso}} = 2.09^{+0.22}_{-0.26} \times 10^{52}$ erg s$^{-1}$, respectively.

In order to reveal the possible physical origin of GRB 201221D, we adopted multi-wavelength criteria (e.g., the Amati relation, $\varepsilon$-parameter, amplitude parameter, local event rate density, luminosity function, and the properties of host galaxy) to diagnose the possible physical origin by comparing with other long- and short-duration GRBs. We find that GRB 201221D is located at the gap between Type II and Type I GRBs in the $E_p$–$E_{\gamma,\text{iso}}$ relation. By adopting the $\varepsilon$-parameter of GRB classification (Lü et al. 2010), one can calculate $\varepsilon \sim 0.034$ which precisely lies on the separation line. We also estimate the local event rate density for the peak bolometric luminosity of ($\sim 2.09 \times 10^{52}$ erg s$^{-1}$) derived from this single event as $\sim 3.0^{+6.9}_{-2.5} \times 10^{-3}$ Gpc$^{-3}$ yr$^{-1}$. The same analysis method proposed in Lü et al. (2014) is used to calculate its amplitude parameter as $f = f_{\text{eff}} \sim 2.13$. Moreover, we also extract the host galaxy information including the stellar mass $M_* = (3.9 \pm 3.1) \times 10^9 M_\odot$, half light radius $R_{50} = 4.4$ kpc of the host galaxy, the normalized offset from the center of the host to the GRB $r_{\text{off}} = R_{\text{off}}/R_{50} = 1.01^{+1.82}_{-0.15}$, and the cumulative light fraction $F_{\text{light}} = 0.38^{+0.44}_{-0.37}$.

Moreover, by extracting the light curve and spectrum of GRB 201221D, we find that $\hat{\alpha}$ is always larger than that of $2 + \hat{\beta}$ during the prompt emission phase. This result strongly suggests that the emission region is undergoing bulk acceleration. By calculating the distance of this region from the central engine, one has $R_{\text{GRB}} = \frac{\Gamma_{\text{jet}}^2 ct}{1+z} \sim 10^{16}$ cm with Lorentz factor $\Gamma_{\text{jet}} \sim 450$ and typical $t \sim 0.1$ s. This distance is not consistent with the predictions of both the photosphere and internal shock models with thermally driven bulk acceleration, but is favored by the ICMART model. If this is the case, the jet is Poynting-flux-dominated and undissipated until reaching a large enough distance (e.g., $\sim 10^{15}$ cm), and the dissipated energy is used to radiate in γ-ray (prompt emission) and accelerate the ejecta.

Combined with the above analysis, the origin of GRB 201221D is favored to be a binary neutron star merger. This model can be used to interpret some observations of this case, except for the $\varepsilon$-parameter and the inconclusiveness of the Amati relation. If this is the case, the remnant of such a merger should leave either a magnetar or black hole which produces a short-duration GRB with a Poynting-flux-dominated jet. It also naturally explains undergoing bulk acceleration in the prompt emission phase. Moreover, another possibility is that GRB 201221D originates in a black hole–neutron star (BH–NS) merger at high redshift. Mandhai et al. (2021) pointed out that the probabilities of BH–NS systems for producing GRBs peak at higher redshifts (e.g., $z > 1$), although production of short-duration GRBs is rarer with BH–NS mergers (Shibata & Taniguchi 2011).

On the other hand, the high-$z$ short-duration GRBs without extended emission are very rare, with only five GRBs (060801, 090426, 111117A, 121226A, and 190627A) observed so far. Within these five GRBs, GRB 111117A has been discussed as having a high probability of originating in a compact binary merger (Dichiara et al. 2021). Moreover, Wiggins et al. (2018) estimated the rate for short GRBs peaks in the redshift range $z = 0.6$–1 by studying the population synthesis, and this is similar to observations of short GRBs with a known redshift distribution (Fong et al. 2015). Also, we expect more and more short GRBs with high-redshift to be observed in the future, and they can be used to study the star formation delay time (Wanderman & Piran 2015). If the binary neutron star mergers indeed occurred in high-$z$ region of early universe, there are important implications for understanding binary stellar evolution, heavy element nucleosynthesis, and chemical evolution, or the possibility of the high-$z$ short-duration GRBs belonging to a different population of bursts (e.g., NS–BH or WD–WD merger systems). However, owing to the relative lack of expected mergers in the high-$z$ region, to understand those implications remains an open question. We hope that there are more neutron star mergers at high redshifts than that expected in the future. The third generation of GW detectors, such as Cosmic Explorer (Reitze et al. 2019) or the Einstein Telescope (Punturo et al. 2010), may play a crucial role in understanding the mergers of these objects out to redshift $z \sim 2$–3.

## Acknowledgments

We acknowledge the use of the public data from the Swift and Fermi data archive. This work is supported by the National Natural Science Foundation of China (grant Nos. 11922301 and 12133003), the Guangxi Science Foundation (grant Nos. 2017GXNSFFA198008 and AD17129006). LHJ thanks support by the Program of Bagui Young Scholars Program, and special funding for Guangxi distinguished professors (Bagui Yingcai & Bagui Xuezhe). B.B.Z. acknowledges support by the National Key Research and Development Programs of China (2018YFA0404204), the National Natural Science Foundation of China (grant Nos. 11833003 and U2038105), and the Program for Innovative Talents, Entrepreneur in Jiangsu. H.S. acknowledges support by the Strategic Priority Research Program of the Chinese Academy of Sciences (grant Nos. XDA15310300, XDA15052100 and XDB23040000). Y.L. acknowledges support by the National Natural Science Foundation of China (grant Nos. 12041306 and 12103089), and the Natural Science Foundation of Jiangsu Province (grant No. BK20211000).